\documentclass[12pt]{article}
\usepackage{amssymb}
\usepackage{graphicx}

\begin{document}

\newcommand \be  {\begin{equation}}
\newcommand \bea {\begin{eqnarray} \nonumber }
\newcommand \ee  {\end{equation}}
\newcommand \eea {\end{eqnarray}}

\title{  Reflection Time Difference as a probe of S-matrix Zeroes in Chaotic Resonance Scattering \footnote{The text is based on
the presentation at the 9th Workshop on Quantum Chaos and Localization Phenomena, May 25--26, 2019, Warsaw, Poland }}

\vskip 0.2cm
\author{Yan V. Fyodorov $^1$ \\
\noindent\small{$^1$ King's College London, Department of Mathematics, London  WC2R 2LS, United Kingdom}}

\maketitle

\begin{abstract}
Motivated by recent interest in the zeroes of $S-$matrix entries in the complex energy plane, I use the Heidelberg model of resonance scattering to introduce the notion of
Reflection  Time Difference which is shown to play the same role for the {\it zeroes} as the  Wigner time delay plays for the $S-$ matrix {\it poles}.

\end{abstract}

\section{Introduction}
The phenomenon of chaotic resonance scattering of quantum waves (or their classical analogues) has attracted considerable theoretical and experimental interest for the last three decades, see e.g. articles in \cite{chao05} and recent reviews \cite{kuhl13,grad14,diet15,hcao15}. The resonances manifest themselves via fluctuating structures in scattering observables, and understanding their statistical properties as completely as possible remains an important task. The main object in such an approach is the energy-dependent $M\times M$ random unitary scattering matrix $S(\lambda)$, $S^{\dagger}(\lambda)S(\lambda)={\bf 1}_M$ which relates amplitudes of incoming and outgoing waves at a given value of the spectral parameter (energy) $\lambda$. Here the integer $M$ stands for the number of open channels at a given value of energy, the dagger denotes the Hermitian conjugation and ${\bf 1}_M$ is the $M\times M$ identity matrix. As is well-known, statistics of fluctuations of the scattering observables over an energy interval comparable with a typical separation $\Delta$ between positions of the resonances can be most successfully achieved in the framework of the so called 'Heidelberg approach' going back to the pioneering work \cite{Ver85}, and reviewed from different perspectives in \cite{MRW2010},  \cite{FSav11} and \cite{Schomerus2015}.
In such an approach the resonance part of the $S$-matrix describing $N\gg M$ interacting resonances is expressed via the Cayley transform in terms of the resolvent of the Hamiltonian $H$
representing the closed counterpart of the scattering system, and which is assumed to be modelled by a $N\times N$  matrix $\mathbf{H}_N$. Namely:
\begin{equation}\label{KW}
\mathbf{S}(\lambda)=\frac{{\bf 1}_M-i\mathbf{K}}{ {\bf 1}_M+i\mathbf{K}},\quad \hbox{with}\quad\mathbf{K}=
\mathbf{W}^{\dagger}\frac{1}{\lambda{\bf 1}_{N}-\mathbf{H}_N}\mathbf{W},
\end{equation}
where $\mathbf{W}$ is the  $N\times M$ matrix  containing the couplings between the channels and the scattering domain. Such couplings are assumed to be energy independent in the relevant energy range, which constitutes the main assumption behind the model. For $\lambda\in\mathbb{R}$, the unitarity of $\mathbf{S}(\lambda)$ follows from Hermiticity of the Hamiltonian $\mathbf{H}_N$. The $M\times M$  matrix $\mathbf{K}$ is known in the literature as the Wigner reaction $K-$ matrix.
To study  fluctuations induced by chaotic wave scattering one then follows the paradigm of relying upon the well-documented random matrix properties of the underlying Hamiltonian operator $H$  describing quantum or wave chaotic behaviour of the closed counterpart of the scattering system. Within that approach one proceeds with replacing $\mathbf{H}_N$ with a random matrix taken from one of the classical ensembles:  Gaussian Unitary Ensemble (GUE, $\beta=2$), if one is interested in the systems with broken time reversal invariance or Gaussian Orthogonal Ensemble (GOE, $\beta=1$), if such invariance is preserved and no further geometric symmetries are present in the system.  The columns $\textbf{w}_a, \, \, a=1,..,M$ of the coupling matrix  $\mathbf{W}$ can be considered either as fixed orthogonal vectors \cite{Ver85} (complex for $\beta=2$ or real for $\beta=1$), or alternatively
 as independent Gaussian-distributed random vectors orthogonal on average \cite{SokZel89}. The results turns out to be largely insensitive to the choice of the coupling as long as inequality $M\ll N\to \infty$ holds in the calculation.
 The approach proved to be extremely successful, and quite a few S-matrix or $K$-matrix characteristics were thoroughly investigated in that framework in the last two decades, either by the variants of the supersymmetry method or related random matrix techniques, see e.g.  \cite{Ver85,Fyo05} as well as more recent results in \cite{Kum13,FN2015,Kum17,FF19}.
 Such calculations are found in general to be in good agreement with available experiments in chaotic electromagnetic resonators (``microwave billiards''), dielectric microcavities and acoustic reverberation cameras ( see reviews \cite{kuhl13,grad14,diet15,hcao15}) as well as with numerical simulations of scattering in such paradigmatic model as quantum chaotic graphs \cite{kott00} and their experimental microwave realizations \cite{microwgraphs1,microwgraphs1a,microwgraphs2,microwgraphs3}.
 It is important however to stress, that theoretical results based on the random matrix description of $\textbf{H}_N$ are {\it universal}, i.e. insensitive to details of the model (and hence  meaningful for description of the experiments), provided the energy scales involved in the processes are comparable with the mean level spacing $\Delta$ in the {\it closed} counterpart of the scattering system, with all channels detached or closed. Only such scales correspond in time-domain to long enough processes of resonance character.

Equivalently, entries $S_{ab}(\lambda)$ of the scattering matrix can be rewritten as \cite{Ver85}
\begin{equation}\label{2}
S_{ab}(\lambda)=\delta_{ab}-2i\, \mathbf{w}^{\dagger}_{a}\left[\frac{1}{\lambda\mathbf{1}_N-{\cal H}_{eff}}\right]\mathbf{w}_{b},
\end{equation}
with an effective non-Hermitian Hamiltonian
\begin{equation}\label{3}
 {\cal H}_{eff}=\mathbf{H}_N-i\mathbf{\Gamma}, \, \quad \, \mathbf{\Gamma}=WW^{\dagger}\ge 0
\end{equation}
whose $N$ complex eigenvalues $\lambda_n=E_n-i\Gamma_n$  provide poles of the scattering matrix in the complex energy plane $\lambda$, commonly referred to as the {\it resonances}\footnote{Note that in some papers the convention $\lambda_n=E_n-i\frac{\Gamma_n}{2}$ is used.} . Note that the scattering matrix of a system without gain or loss is unitary, and necessarily has all poles in the lower
half of the complex energy plane $Im({\cal \lambda}) < 0$. Such poles are accompanied by the corresponding zeroes in the upper half ($Im({\cal \lambda}) > 0$),
whose positions are exact mirror images of the poles in  the real axis.  This fact is most clearly illustrated by the relation
\be\label{phaseshift}
\det(\hat{S}(\lambda))=\frac{\det\left(\lambda\mathbf{1}_N-\textbf{H}_N-i\mathbf{\Gamma}\right)}
{\det\left(\lambda\mathbf{1}_N-\mathbf{H}_N+i\mathbf{\Gamma}\right)}:=e^{i\delta(\lambda)}
\ee
where real $\delta(\lambda)$ is known as the scattering phaseshift.
The statistics of real positions $E_n$ and widths $\Gamma_n$ of resonant poles, as well as statistics of the associated residues related to non-orthogonal eigenvectors of non-Hermitian Hamiltonian ${\cal H}_{eff}$
has been subject of considerable research activity in this framework for a few decades theoretically \cite{SokZel89,haak92,lehm95a,fyod96,Fyo97a,fyod99,somm99,jani99,scho00,sant01,fyod02,Kott05,poli09,Celardo11,SZel12,FyoSav12,fyod15,Lippolis,kozh17}
and in the last decade became also accessible experimentally \cite{kuhl08,koeh10,difa12,bark13,liu14,gros14,DavyGenack18,DavyGenack19}.
At the same time,
as due to unitarity all information about zeroes is redundant, those hardly attracted any attention until recently.

The situation changed with the recent advent of the
 phenomenon of coherent perfect absorption (CPA)\cite{CPA}. This phenomenon is also commonly referred to as ‘anti-lasing’
because it corresponds in a certain sense to reversing the process of coherent
emission of radiation at the lasing threshold, with the goal being to ensure that there is no
outgoing waves for some nonzero input into the system.
The particular challenge is to realize such effect in chaotic or disordered
media without special symmetries.

To this end it was suggested in \cite{Li2017,FST2018} that CPA may be engineered by adding {\it spatially-localized} losses making the
scattering system non-flux-conserving/nonunitary. Such operation moves S-matrix zeroes and poles around in
the complex energy plane so that they lose a simple relation to each other. To illustrate this in the framework of the Heidelberg approach let us consider a single point-like local absorber and treat it as additional unobservable {\it absorbing channel}. Such a channel can be most naturally characterized by a  $N-$vector  ${\cal A}$ with the norm $|{\cal A}|^2=\gamma_0>0$ assumed to be orthogonal to
 the space spanned by all scattering channel vectors $\textbf{w}_{a}, a=1,\dots M$. The parameter $\gamma_0$ characterizes the
  strength of the absorber, i.e. the amount of flux lost irretrievably through this new channel.
  As the result, for any $\gamma_0\ne 0$ the former unitary $M\times M$ scattering matrix $S(\lambda)$ becomes subunitary and can be most easily understood as   $M\times M$ diagonal sub-block $S(\lambda,\gamma_0)$ of the full unitary $(M+1)\times (M+1)$
scattering matrix ${\cal S}$.  Introducing the rank-one $N\times N$ matrix $\mathbf{\Gamma}_A={\cal A}{\cal A}^{\dagger}\ge 0$
it is easy to show \cite{fyod04} that such block $S(E,\gamma_0)$ can be written in the form
generalizing (\ref{KW}):
\be\label{KWA}
S(\lambda,\gamma_0)=\frac{\mathbf{1}-i\mathbf{K}_A}{\mathbf{1}+i\mathbf{K}_A},\quad \mathbf{K}_A=\mathbf{W}^{\dagger}\frac{1}{\lambda\mathbf{1}_N-\mathbf{H}_N+i\mathbf{\Gamma}_{A}}\mathbf{W}
\ee
which further implies
\be
\det{S}(\lambda,\gamma_0)=\frac{\det{\left(\lambda-\mathbf{H}_N+i\mathbf{\Gamma}_{A}-i\mathbf{\Gamma}\right)}}
{\det{\left(\lambda-\mathbf{H}_N+i\mathbf{\Gamma}_{A}+i\mathbf{\Gamma}\right)}}
\ee
showing that now the poles of $S(\lambda,\gamma_0)$ are given by the $N$ complex eigenvalues of
the effective non-Hermitian Hamiltonian ${\cal H}^{(+)}_{eff}=\mathbf{H}_N-i\left(\mathbf{\Gamma}+\mathbf{\Gamma}_{A}\right)$
whereas  zeroes of $S(\lambda,\gamma_0)$ are given by the $N$ complex eigenvalues of a different
 effective non-Hermitian Hamiltonian:
\be\label{Nonhermabsorp}
 {\cal H}^{(-)}_{eff}=\mathbf{H}_N+i\left(\mathbf{\Gamma}-\mathbf{\Gamma}_{A}\right)
\ee
Hence zeroes are no longer related to poles as long as the absorber is present: ${\cal A}{\cal A}^{\dagger}\ne 0$.
In particular, it is easy to understand that the system turns into a random anti-laser at exactly
those energies and loss values at which any of the $S$-matrix zeroes
crosses the real energy axis. And indeed, such CPA has been recently experimentally realized in
an elegant experiment \cite{antilasing}.

These developments naturally attracted interest to S-matrix zeroes which can now be located in the whole complex plane.
Their properties clearly deserve to be studied, both theoretically and experimentally.
Recently the density of those zeroes in the complex plane  has been efficiently described in the RMT framework, both perturbatively for weak losses \cite{Li2017} and non-perturbatively for arbitrary losses \cite{FST2018}.  An important question then
arises how the positions of the complex $S-$matrix zeroes can be extracted from scattering measurements.

To this end let us first recall that the S-matrix poles manifest themselves in a most explicit way via the so-called {\it Wigner time delay} $\tau_W(\lambda)$ defined via the energy derivative of the scattering phaseshift per scattering channel \cite{wigner55}. As follows from (\ref{phaseshift}) in the present model Wigner time delay is exactly given by the sum of Lorentzians of widths $\Gamma_n$ centered at positions $E_n$,
\begin{equation}\label{wigner}
\tau_W(\lambda):=\frac{1}{M}\frac{d}{d\lambda}\delta(\lambda)=-i\frac{d}{d\lambda}\ln{\det S(\lambda)}=\frac{2}{M}\sum_{n=1}^N\frac{\Gamma_n}{(\lambda-E_n)^2+\Gamma_n^2}
\end{equation}
Statistics of Wigner time delays and related quantities in wave-chaotic scattering attracted a considerable interest for about three decades and is still an active research topic, see  \cite{Lehmann95,FSS96,FSS97,Brower99,SFS01,OssFyo05,MezSimm13,Novaes15,Cunden15,Texier16,Grabsch18,Oss18} and references therein.
Interestingly, $\tau_W(\lambda)$ naturally appears in characterization of systems with {\it spatially-uniform} absorption.
In fact such absorption is always present in any realistic system, e.g. due to losses in resonator walls in microwave cavities.
It can be very simply accounted for by allowing the spectral parameter $\lambda$ to have a small imaginary part $\epsilon=Im (\lambda)>0$.  It is easy to see that such a replacement makes the full scattering matrix subunitary: $|\det S(\lambda+i\epsilon)|<1$. Moreover, expanding eq.(\ref{phaseshift}) for weak enough uniform losses $\epsilon \ll \Delta$  shows that
   the degree of non-unitarity is exactly controlled by the Wigner time-delay as
\begin{equation} \label{subunit} |S(\lambda+i\epsilon)| \approx e^{-\epsilon \, M\tau_W(\lambda)}
\end{equation}
 This fact has been recognized long ago in an early paper \cite{Doron1990} and the relation (\ref{subunit})  used there to effciently extract Wigner time delays from unitary deficit of experimentally measured scattering in a single channel $M=1$ wave-chaotic system.

 To see how a similar construction may help to get access to $S-$matrix zeroes, we consider
 a guiding example of a {\it flux-conserving} two-channel time-reversal invariant scattering system characterized by the unitary $2\times 2$ scattering matrix which we write as $ \hat{S}(\lambda)=\left(\begin{array}{cc}R_1(\lambda) & t(\lambda)\\ t(\lambda) & R_2(\lambda) \end{array}\right)$. Flux conservation implies unitarity which then imposes the following relations between complex entries of the matrix (suppressing the argument $\lambda$ for simplicity):
 \be
 R_{1,2}=|R|e^{i\phi_{\tiny{1,2}}}, \quad t=|t|e^{i\theta}, \quad e^{2i\theta}=-e^{i\left(\phi_{1}+\phi_{2}\right)}
 \ee
On the other hand, $R_{1,2}$ can be easily represented in the Heidelberg approach as (cf. (\ref{KWA}))
\be\label{KW2}
R_{1,2}(\lambda)=\frac{1-iK_{1,2}}{1+iK_{1,2}},\quad K_1=\mathbf{w}_1^{\dagger}\frac{1}{\lambda\mathbf{1}_N-\mathbf{H}_N+i\mathbf{\Gamma}_2}\mathbf{w}_1, \,\,
K_2=\mathbf{w}_2^{\dagger}\frac{1}{\lambda\mathbf{1}_N-\mathbf{H}+i\mathbf{\Gamma}_1}\mathbf{w}_2
\ee
where as before we used the notations: $\mathbf{\Gamma}_{1}=\mathbf{w}_{1}\otimes \mathbf{w}_1^{\dagger}$ and
$\mathbf{\Gamma}_{2}=\mathbf{w}_{2}\otimes \mathbf{w}_2^{\dagger}$.
 By using straightforward algebraic manipulations it is then easy to show that
\be
\frac{R_1(\lambda)}{R_2(\lambda)}=e^{i\left(\phi_{1}-\phi_{2}\right)}=\frac{\det{\left(\lambda\mathbf{1}_N-\mathbf{H}_N-
i\mathbf{\Gamma}_{1}+i\mathbf{\Gamma}_2\right)}}
{\det{\left(\lambda\mathbf{1}_N-\mathbf{H}_N+i\mathbf{\Gamma}_{1}-i\mathbf{\Gamma}_2\right)}}
\ee
Thus for the flux-conserving scattering system the ratio $R_1(\lambda)/R_2(\lambda)$ is unimodular, and its poles in the complex $\lambda-$plane are
eigenvalues of the effective Hamiltonian ${\cal H}^{(-)}_{eff}=\mathbf{H}_N+i\left(\mathbf{\Gamma}_1-\mathbf{\Gamma}_{2}\right)$.
The latter  after straightforward re-interpretation is nothing else but  exactly the same object
as one featured in our description of the CPA, see (\ref{Nonhermabsorp}) for the particular choice $M=1$.
Moreover, motivated by  Eq.(\ref{wigner})
one may introduce the notion of {\it Reflection Time Differences} via the associated energy derivatives
\begin{equation}\label{RTDdef}
\delta {\cal T}(\lambda):=\frac{1}{2}\frac{\partial }{\partial \lambda}\left(\phi_1-\phi_2\right)=-\frac{i}{2}\frac{\partial}{\partial \lambda}\ln{\left(\frac{R_1}{R_2}\right)}=
\sum_{n=1}^N \frac{\Im {\cal Z}_n}{\left(\lambda-\Re {\cal Z}_n\right)^2+\left(\Im {\cal Z}_n\right)^2},
\end{equation}
with ${\cal Z}_n=\Re {\cal Z}_n+i\Im {\cal Z}_n$ being  exactly the complex zeroes of $R_1$.
This construction makes it obvious that $\delta {\cal T}(\lambda)$ plays for the $S-$matrix {\it zeroes} the same role as the  Wigner time delay plays for the $S-$ matrix {\it poles}. The only essential difference is that due to presence of zeroes in both halves of the complex plane each Lorentzian term enters in the sum in (\ref{RTDdef}) with its own sign, which can be either positive or negative.

In particular, consider an interval of (real) energies $[\lambda_1,\lambda_2]$ such that $|\lambda_1-\lambda_2|\gg \Delta$. Then
\be
\int_{\lambda_1}^{\lambda_2}\delta{\cal T}(\lambda) \, d\lambda={\pi}\left({\cal N}_{+}- {\cal N}_{-}\right)
\ee
where ${\cal N}_{+}$ and ${\cal N}_{-}$  are numbers of complex zeroes whose real parts $\Re {\cal Z}_n\in [\lambda_1,\lambda_2]$ and imaginary
parts $\Im {\cal Z}_n$ are respectively positive/negative. Moreover, incorporating now into our model weak uniform losses $\epsilon \ll \Delta$ one has the relation:
\begin{equation} \label{subunitR}
\left|\frac{R_1}{R_2}\right|_{\lambda+i\epsilon} \approx e^{-2\epsilon \, \delta{\cal T}(\lambda)}
\end{equation}
fully analogous to (\ref{subunit}) for the Wigner time delay. Such a relation clearly provides a possibility
 to extract $\delta{\cal T}(\lambda)$ from experimental measurements of the unitary deficit of the ratio of reflection coefficients in weakly absorbing microwave cavities with two open channels. Extension to arbitrary number of open channels is very straightforward, and basically amounts to using
 $\det{R_1}$ and $\det{R_2}$ instead of $R_1$ and $R_2$ in the definition of  $\delta{\cal T}(\lambda)$.

  Finally, the use of the reflection time difference $\delta{\cal T}(\lambda)$ opens a possibility to extract positions of complex zeroes
from the sum of (sign-weighted) Lorentzians in (\ref{RTDdef}), e.g by using the {\it harmonic inversion} method, see \cite{kuhl08} and refs. therein.
 To test this possibility in principle,  a particular realization of the Heidelberg model corresponding to $N=400$ zeroes has been numerically generated. Two open channels were chosen with particular values of coupling parameters $\gamma_{1,2}=\mathbf{w}^{\dagger}_{1,2}\mathbf{w}_{1,2}$, and the poles were extracted by employing harmonic inversion from  the signal $\delta{\cal T}(\lambda)$ generatated in two different ways: via the energy derivative according to (\ref{RTDdef}) and from the unitary deficit in presence of a small uniform absorption as in (\ref{subunitR}). The extracted zeroes were compared with their true values obtained by the direct diagonalization of the associated non-Hermitian Hamiltonian ${\cal H}^{(-)}_{eff}$. The results are presented in the figure below.

\begin{figure}[!htb]
\includegraphics[width=0.5\columnwidth]{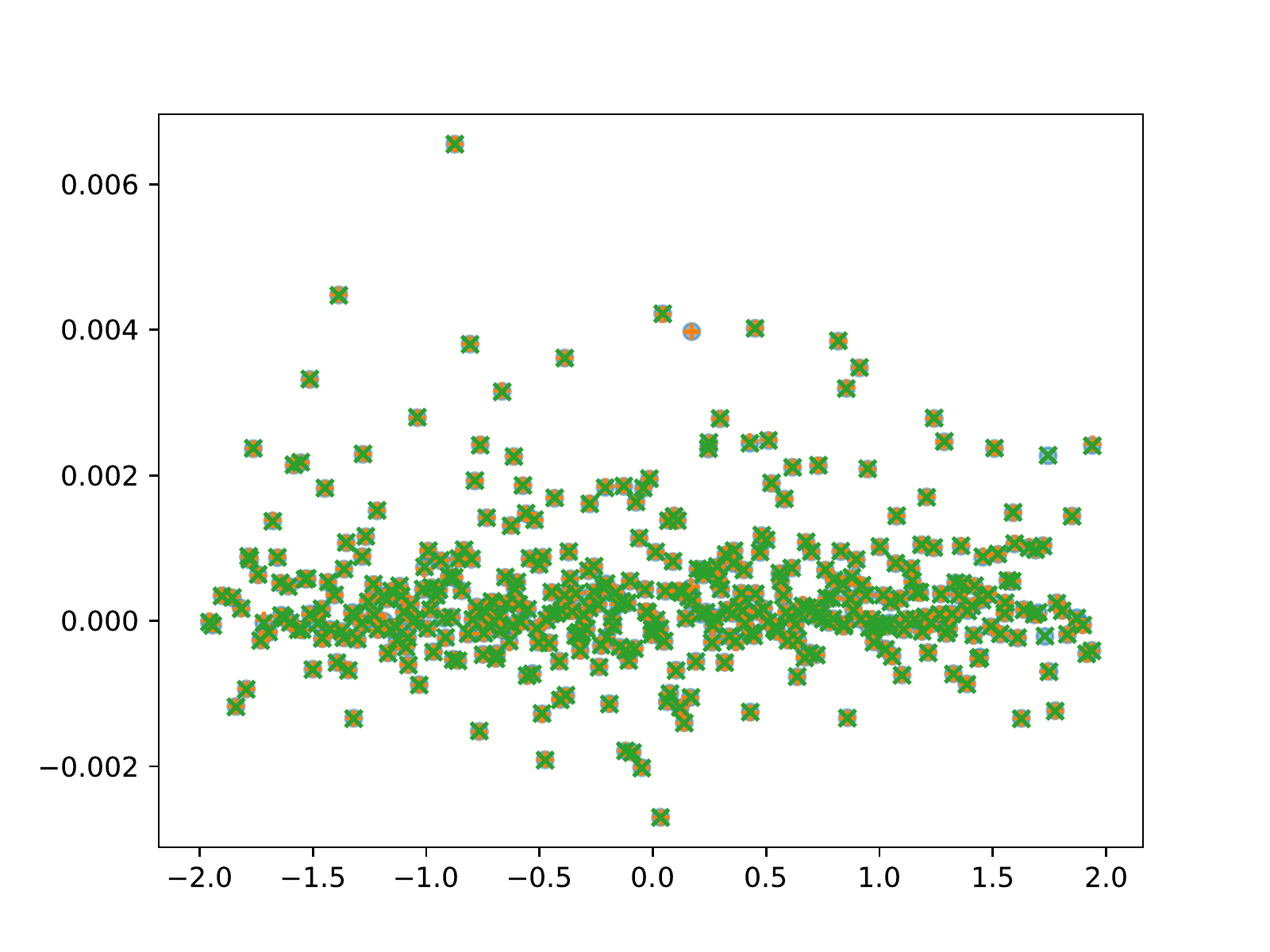}\\
 {\small Fig.1: \sf RMT simulations of two-channel Heidelberg model with $N=400$, $\gamma_1=0.1$,   $\gamma_2=0.05$, $\epsilon=10^{-5}$.
Blue circles are true zeroes, red $+$ signs are extracted using Eq.(\ref{RTDdef}), and red $\times$ signs using Eq.(\ref{subunitR}).
{\it Courtesy of  M. Osman.}}
\end{figure}

In conclusion, in this paper I suggest to use the Reflection Time Difference  $\delta{\cal T}(\lambda)$ as a probe of S-matrix  zeroes
in the complex energy plane. It would be certainly interesting to develop RMT approach further in order to characterize the statistical properties of $\delta{\cal T}(\lambda)$ in systems with chaotic wave scattering (such as distributions, correlation functions, etc.).  This task as well as experimental studies of both $\delta{\cal T}(\lambda)$ and the associated $S-$matrix zeroes for real systems
remain among the outstanding challenges.

{\bf Acknowledgments:} The author is grateful to Mr. Mohammed Osman for performing RMT simulations and producing the figure for this article.
This research has been supported by the EPSRC grant EP/N009436/1 "The many faces of random characteristic polynomials".

\end{document}